\documentclass[pre,twocolumn,superscriptaddress,floatfix]{revtex4-2}
\usepackage{dcolumn}
\usepackage{bm}
\usepackage{amsmath}
\usepackage{amssymb}
\usepackage[english]{babel}
\usepackage{graphicx}

\DeclareMathOperator{\Ei}{Ei}

\begin{document}

	\preprint{APS/123-QED}
	
\title{Conductivity of concentrated salt solutions}

\author{Olga I. Vinogradova}
\email[Corresponding author: ]{oivinograd@yahoo.com}
\affiliation{Frumkin Institute of Physical Chemistry and Electrochemistry, Russian Academy of Sciences, 31 Leninsky Prospect, 119071 Moscow, Russia}

\author{Elena F. Silkina}

\affiliation{Frumkin Institute of Physical Chemistry and Electrochemistry, Russian Academy of Sciences, 31 Leninsky Prospect, 119071 Moscow, Russia}
\begin{abstract}
The conductivity of concentrated salt solutions has posed a real puzzle for theories of electrolytes. Despite a quantitative understanding of dilute solutions, an  analytical theory for concentrated ones remains a  challenge for almost a century, although a number of parameters and effects incorporated into theories increases with time. Here we show that the conductivity of univalent salt solutions can be perfectly interpreted using a simplest model that relies on
a modified mean-field description of electrostatic
interactions and
on a classical approach to calculating  colloid electrophoresis. We derive a compact equation, which predicts that the ratio of conductivity to that at an infinite dilution is the same for all salt and depends only on a product of the harmonic mean of ion hydrodynamic radii and the square root of concentration. Our equation fits very well the data for inorganic salts (up to a few mol/l), although at a very high dilution the relaxation correction seems necessary.
\end{abstract}

	\date{\today}
	\maketitle

\section{Introduction}\label{sec:intro}
The electrolyte solutions are ionic conductors,  thanks to cations and anions formed as a result of dissociation. The electrical conductivity $K$ is one of their most important trait, which is widely used in chemical, biological as well as other applications, and the performance of many (\emph{e.g.} energy storage) devices depends entirely on the ionic transport~\cite{a.dediego:2001,thirstrup.c:2021,gu.y:2023}. The amount of water in Earth's mantle is inferred from the conductivity data~\cite{yoshino.t:2013}. Besides, historically, the conductivity is the most important source of information on electrolyte properties (\emph{e.g.} ion pairing)~\cite{marcus.y:2006}.

The physical origin of  ionic conductivity is more or less clear. If a direct electric field $E$ is applied, ions migrate relative to a solvent with a constant speed towards one of the electrodes. Such a migration results in an electric current, which density obeys Ohm's law, $J = KE$, provided the field is weak enough. The speeds of ions are given by
\begin{equation}\label{eq:V_ions}
 V_{\pm} = M_{\pm} E,
\end{equation}
where $M_{\pm}$ are their  mobilities. The current densities induced in an univalent electrolyte are then $J_{\pm} =  e n_{\infty} V_{\pm}$, where $e$ is an elementary positive charge and $n_{\infty}=n_{\pm}$ is the number density (concentration) of an electrolyte solution. Consequently, the current density of a solution reads $J = J_+ - J_- = e E n_{\infty} M,$
where $M = M_+ - M_-$ is the difference in mobilities of cations and anions,
and
\begin{equation}
 K  = e n_{\infty} M
\end{equation}
Thus, the calculations of the conductivity at a given concentration are reduced to those of $M$.

The quantitative understanding of ion mobilities is a challenging  problem that has been addressed over nearly a century and by many groups, which is often termed a central issue
of chemical physics~\cite{bernard.o:2023,naseri.b:2023}. The simplest expression for univalent ions can be derived by postulating the Stokes resistance to the ion propulsion
\begin{equation}\label{eq:inf}
M_{\pm} \simeq M_{\pm}^0 \simeq \pm\dfrac{e }{6\pi \eta R_{\pm}},
\end{equation}
where $\eta$ is the dynamic viscosity of the solvent and $R_{\pm}$ are the hydrodynamic radii of the cation and anion. $M_{\pm}^0$ given by Eq.\eqref{eq:inf} and termed the mobilities at an \emph{infinite dilution} do not depend on salt concentration. In this model
\begin{equation}\label{eq:Kinf}
M \simeq M_0 \simeq \dfrac{e }{3 \pi \eta R_{h}} \,  \mathrm{and} \, K \simeq K_0 \simeq \dfrac{e^2 n_{\infty }}{3 \pi \eta R_h},
\end{equation}
where $\eta$ is the dynamic viscosity of a solvent and $R_h = 2 (R_+ R_-)/(R_+ + R_-)$ is the harmonic mean of hydrodynamic radii. However, experiments on a conductivity show that $M$ and $K$ are generally smaller than predicted by \eqref{eq:Kinf} and this discrepancy augments on increasing salt concentration~\cite{kohlrausch.f:1900,a.dediego:2001}. Chemists  have long (since Arrhenius) used this fact to infer a degree of dissociation (or an ion pairing). Physicists might  view  this simply as  unreliability of hydrodynamic arguments based on the Stokes drag force.

There is a large literature describing attempts to provide a satisfactory theory of electrolyte conductivity.
One of the first systematic treatments of the influence of $n_{\infty }$ on conductivity was contained in a remarkable paper by~\citet{onsager.l:1927} who argued that    the relative conductivity reduction $K/K_0$ is attributed to two contributions. One (and the main) is ionic electrophoresis, which is the migration of ion surrounded by  equilibrium screening cloud. Second is a so-called relaxation correction that depends on the cloud distortion produced by the motion in the external field. Onsager showed that at a \emph{very high dilution} $K/K_0$ decreases linearly with the square root of the concentration. In efforts to better understand the connection between the ion mobilities and salt concentration  many authors extended the theory, but failed to come to grips with the conductivity at high salt~\cite{fuoss.rm:1978}.
However, most electrolyte solutions in nature and various applications
are concentrated. Thus, salt concentration in human blood plasma is about  0.15 mol/l, in the Atlantic Ocean it is  $\simeq 0.6$ mol/l, Li-ion  batteries and supercapacitor electrolytes are usually of concentration $1-2$ mol/l, reference electrolytes of pH-meters and glass micropipette electrodes - of concentration 3 mol/l.

 Recent attempts at improvements on the theory have been focused on a sophisticated description of so-called electrostatic and hydrodynamic effects based on microscopic concepts. Various techniques, such as the mean spherical approximation~\cite{dufreche.jf:2005,bernard.o:2023}, mode-coupling theory~\cite{chandra.a:1999,aburto.cc:2013}, density functional approach~\cite{chandra.a:1999,avni.y:2022}, and more~\cite{goldsack.de:1976,villullas.h:2005,banerjee.p:2019,naseri.b:2023,kalikin.nn:2024} have been employed. These publications involve additional parameters and contributions, and mostly rely on numerical calculations.
 This makes them difficult to use and limits prediction capabilities, but simple analytical formulas that apply for concentrated solutions appear to be missing, although there have been some attempts to deduce them from microscopic  theories. A useful analytical solution is known~\cite{bernard.o:1992}, but expressed in terms of special functions and can hardly be called simple or compact. A more recent approximate expression~\cite{avni.y:2022} is inapplicable at molar concentrations due to non physical behavior of both electrophoretic and relaxation terms~\cite{vinogradova.oi:2024}. Thus, in spite of importance for many fields, a theory that has the merit  of yielding useful (approximate) analytical result for the conductivity of concentrated salt solutions as well as makes strong appeal to macroscopic arguments is still absent.

We also gained the impression that researchers in one branch appear to be unaware of closely related work in other branches of physics and chemistry. Indeed, most conductivity publications have either ignored or failed to make direct connection with the theory of electrophoresis developed in colloid science and fluid mechanics. Thus, already in 1921 Smoluchowski~\cite{smoluchowski.m:1921} made clear that the electrophoretic mobility is proportional to the electrostatic surface potential $\Phi_s$, and not, as one might imagine upon first thought, to the charge of the particle. Since in all existing theories ions  are  idealized as hard spheres of radius $a_{\pm}$,  this conclusion holds for ions too, at least within the framework of such a colloid representation.  However, the significance of the Smoluchowski paper does not seem to be recognized in the modern conductivity literature. Recently  we have drawn attention to this drawback of the conductivity theories   and have carried calculations of $\Phi_s$ and then of the so-called zeta potential of ions $Z \propto \Phi_s$, which defines their electrokinetic mobility~\cite{vinogradova.oi:2023b}. Assuming  cations and anions of equal hydrodynamic radius, $R_+ = R_-$, and using $a=R_{\pm}$, we  have derived an equation for $K/K_0$, that applies even at a  \emph{very low dilution} and is more accurate and compact than prior approximate expression~\cite{avni.y:2022}. However, our paper has not given attention to the relaxation issue, which is traditionally invoked in the conductivity theories. It is also, in practice, unlikely that both cations and anions have exactly the same size, so the status of the theory for real salt solutions still remains somewhat obscure.

The present paper concerns with solutions composed of inorganic ions of unequal radius $R_+ \neq R_-$. We revisit the issue of calculation of ion zeta potentials and propose an even much simpler expression than derived before~\cite{vinogradova.oi:2023b}, which appears to be more accurate and applies at a larger concentration range.
We then use it to derive the simplest conductivity  equation valid up to several mol/l, and argue that relative conductivities, $K/K_0$, of all inorganic salts plotted against $R_h \sqrt{n_{\infty}}$ would collapse into a single  curve. This conclusion is supported by providing a comparison with data for several standard salts, which shows that our theory is in excellent agreement with experiment up to concentrations of a few molars. We also demonstrate that although the relaxation effect affects a small decrement to a relative conductivity at high dilution, it becomes negligibly small in concentrated solutions, where the reduction in $K/K_0$ is significant.

\section{General considerations}\label{sec:model}

\begin{figure}[h]
   	\begin{center}
   		\includegraphics[width=0.9\columnwidth]{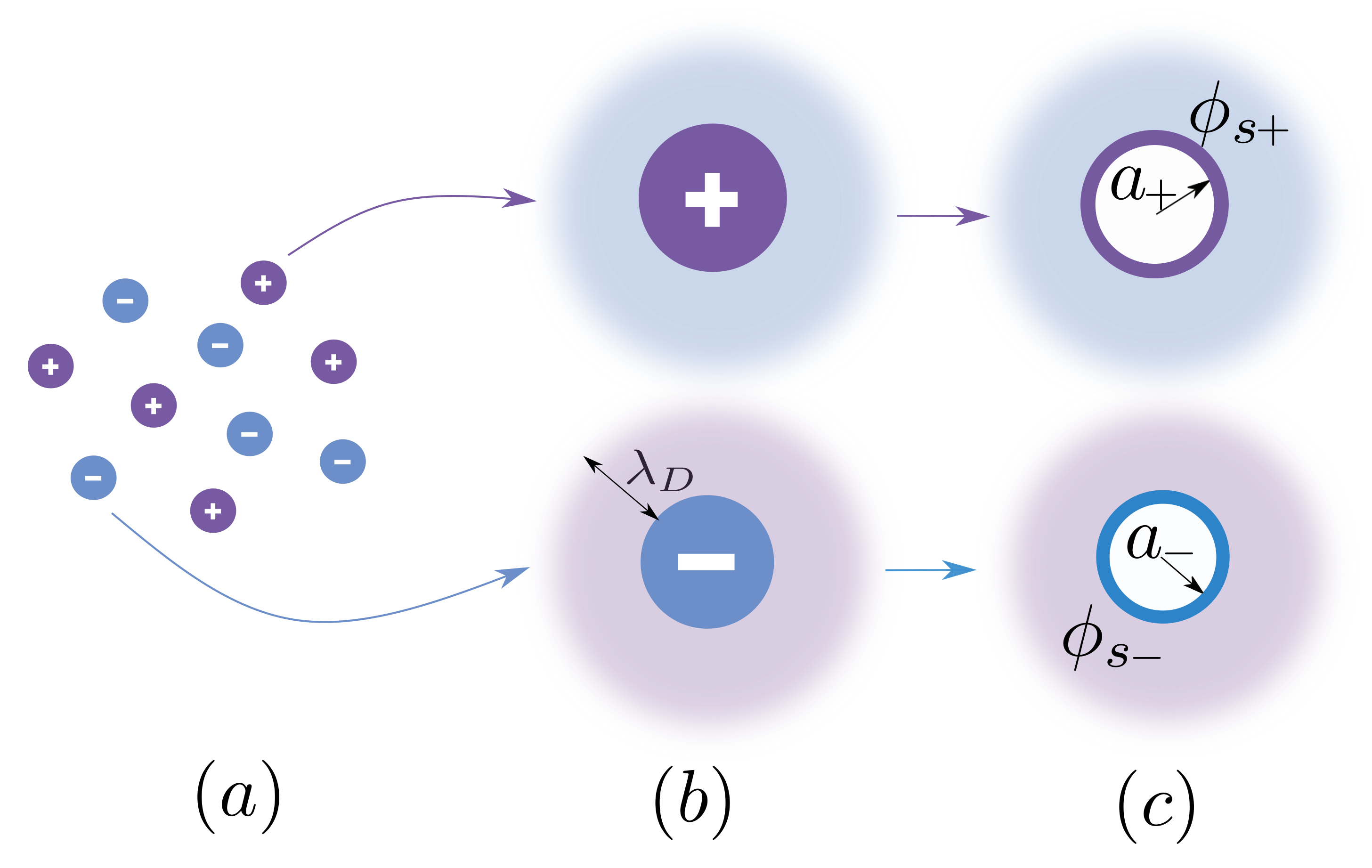}\\
   	\end{center}
   	\caption{Sketch of the charge distribution in the (a) real electrolyte solution, (b) Debye-Hückel classical model of  finite size ions with the charges located at their center, and (c) model of ions with neutral interior, but constant surface charge densities.  }
   	\label{fig:static}
   \end{figure}

We consider a bulk 1:1 salt solution of concentration $n_{\infty} [\rm{m^{-3}}]$ assuming that the description of its global \emph{static} properties can be restricted to a mean-field theory. Our approach is generally based on the fundamental ideas of Debye and Hückel~\cite{debye.p:1923}. Namely, an electrolyte solution with randomly distributed ions [shown in Fig.~\ref{fig:static}(a)] is idealized as a colloidal suspension of so-called central (positive and negative) ions representing nonconducting hard spheres of finite radii $a_{\pm}$ carrying unit charges $\pm e$ at their center. Compensating charges of the opposite sign and equal magnitude stay in the neighborhood of the central ions. These are  the so-called ionic clouds or electrostatic diffuse layers (EDLs).
EDLs are treated as composed by point-like ionic species, and their extension is of the order of the Debye screening length of a solution, $\lambda_{D}=\left( 8\pi \ell _{B}n_{\infty}\right) ^{-1/2}$, where $\ell _{B}=\dfrac{e^{2}}{\varepsilon k_{B}T}$ the Bjerrum length of the solvent, $\varepsilon$ is the solvent permittivity, $k_{B}$ is the Boltzmann
constant, and $T$ is a temperature. This classical model is illustrated in Fig.~\ref{fig:static}(b).

Note that by analyzing the experimental data it is more convenient to use the concentration $c_{\infty}[\rm{mol/l}]$, which is related to $n_{\infty}$ as $n_{\infty} \simeq N_A \times 10^3 \times c_{\infty}$, where $N_A$ is Avogadro's number. The Bjerrum
length of water at $T \simeq 298$~K is equal to about $0.7$ nm leading to a
useful formula for 1:1 electrolyte
\begin{equation}\label{eq:DLength}
  \lambda_D [\rm{nm}] \simeq \frac{0.305 [\rm{nm}]}{\sqrt{c_{\infty}[\rm{mol/l}]} }
\end{equation}
Thus upon increasing $c_{\infty}$ from $10^{-5}$ to $4$ mol/l the screening length (and the cloud extension) is reduced from about 100 down to 0.15 nm.

Some further comments should be made. The fundamental insight of the Debye and Hückel approach is to realize that although the average potential inside the electrolyte solution is zero, there are strong positional correlations between the oppositely charged central ions~\cite{levin.y:2002}. Indeed, if some cation is selected as central, then the position of a central anion cannot be arbitrary anymore, and their local ordering inevitably occurs, although on average they are randomly distributed. Another important  insight is to realize what we would term the point-particle duality of ions (by analogy
to the familiar concept of wave-particle duality of
electrons in quantum mechanics): ions can be either point-like or exhibit properties of the particles of finite size
depending on the circumstances.

This classical model that is widely used to interpret equilibrium properties of electrolyte solutions  will provide the starting point for our discussion of migration of central ions under an applied electric field.

In the interior of the ions of radii $a_{\pm}$ there are no other charges except for the fixed at the centers, and the inner electrostatic potentials satisfy the Laplace equation, $\nabla^2 \phi_{\pm} = 0$, where  $\phi_{\pm}=e\Phi_{\pm}/(k_{B}T)$ are the dimensionless electrostatic potentials. This inner field does not impact the electrophoresis, which depends only on the surface potentials $\phi_{s\pm}$ with respect to the solution far from them (with the zero average potential), as stressed in Sec.~\ref{sec:intro}.
The distribution of electrostatic potentials around the central ions
can be obtained by application of the non-linear Poisson-Boltzmann equation~\cite{joly.l:2024}:
\begin{equation}
\nabla^2 \phi_{\pm} = \dfrac{1}{r^2} \dfrac{d}{d r} \left( r^2 \dfrac{d \phi_{\pm}}{d r} \right)  = \frac{\sinh \phi_{\pm}}{\lambda
_{D}^{2} },   \label{eq:NLPB}
\end{equation}
where $r$ is the distance from the centers of the corresponding spheres.
The  outer potentials would be the same  as for the spheres with a constant surface charge density $\pm e/(4 \pi a_{\pm}^2)$.
and decay from $\phi_{s\pm}$ to zero as $r \to \infty$, which is the first boundary condition that should be imposed to integrate \eqref{eq:NLPB}. While this condition is applied at infinity, as we already discussed above, in reality the EDLs extend only to distances of the order of $\lambda_D$, and thus potentials in fact vanish beyond the clouds.
   The  surface potentials are established self-consistently and  salt-dependent.
 It is therefore admissible for electrophoretic calculations to replace the standard classical model shown in Fig.~\ref{fig:static}(b) by spheres with a constant densities of a surface charge and neutral interior [see Fig.~\ref{fig:static}(c)].

\begin{table}[h]
	  \caption{Some typical values of hydrodynamic radii of ions~\cite{kadhim.mj:2020}, they harmonic means and ratios.}
	\label{table:radii}
	\begin{tabular}{|c|c|c|c|c|}
		 \hline
	Salt &	$R_{+}$, nm & $R_{-}$, nm & $R_{h}$, nm  & $N$  \\
 \hline	KBr &	0.1295        & 0.1179 & 0.123 &  1.1     \\
 \hline	NaCl &  0.184     & 0.1245   & 0.148 &   1.48 \\
 \hline	LiI &	0.238        & 0.1135  & 0.154  & 2.1 \\
  \hline
	\end{tabular}
\end{table}

What is still missing and unknown are the radii $a_{\pm}$, which are the cut-off distances from the selected point charges, where there are no other ions. Some approaches to quantify them have been proposed~\cite{avni.y:2022,bernard.o:2023}, but nevertheless, not developed enough to describe accurately the conductivity data. It is, however, well understood that to describe the \emph{dynamic} response to an external field, the ion hydrodynamic radii $R_{\pm}$ should be taken into account. Inorganic ions have hydrodynamic radii from 0.1 to 0.3 nm~\cite{kadhim.mj:2020}, and we present their values for some univalent electrolytes in Table~\ref{table:radii}, together with harmonic means $R_h = 2 R_+ R_-/(R_++R_-)$ and $N=R_+/R_-$. Since the introduction of $R_{\pm}$ inevitably prevents point-like EDL ions from penetrating inside the sphere, the need to invoke an additional electrostatic cut-off~\cite{avni.y:2022}  is removed. So, below we simply postulate $a _{\pm} = R_{\pm}$ and impose the electrostatic boundary condition $\phi'_{\pm}(r=R_{\pm}) = \mp \ell _{B}/R_{\pm}^2$ to integrate Eq.~\eqref{eq:NLPB} at this surface.

Since the electrostatics of ions involves two length scales, it is convenient to introduce their dimensionless radii  $\varrho_{\pm} = R_{\pm}/\lambda_D$. Note that in the range of $c_{\infty}$ below 2 mol/l the values of $\varrho_{\pm}$ remain smaller than unity or very close to it. Say, if we set $c_{\infty} = 2$ mol/l, then for the largest ion (Li$^+$) in  Table~\ref{table:radii} we obtain $\varrho_+ \simeq 1.1$, but for the smallest (I$^-$) we get $\varrho_- \simeq 0.5$. However, when $c_{\infty} = 4$ mol/l the same ions give $\varrho_+ \simeq 1.6$ and  $\varrho_- \simeq 0.7$.

\begin{figure}[h]
   	\begin{center}
   		\includegraphics[width=0.8\columnwidth]{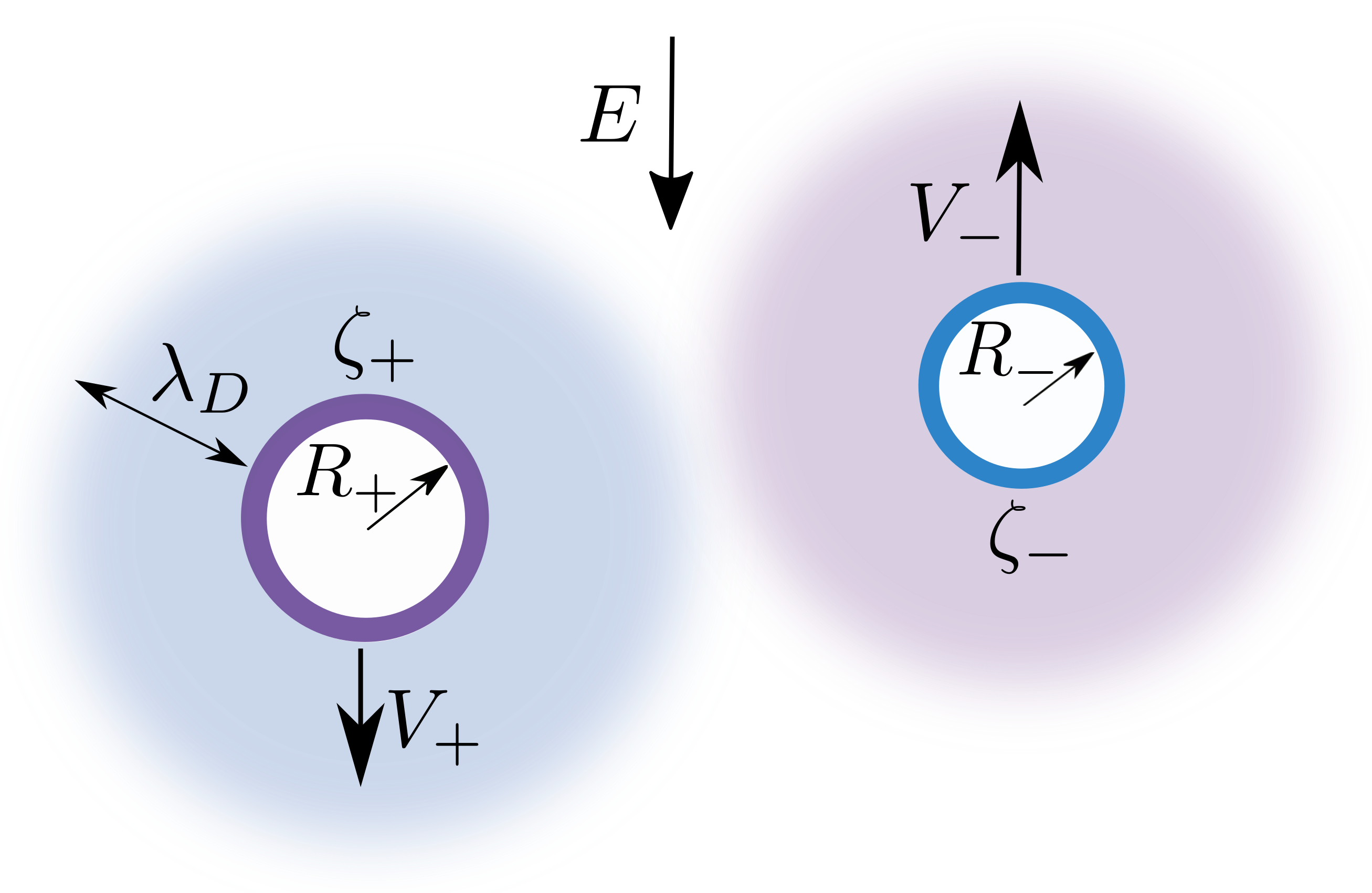}\\
   	\end{center}
   	\caption{Sketch of the cation and anion with hydrodynamic radii $R_+$ and $R_-$ in the bulk electrolyte solution characterized by  $\lambda_D$. The cation propels with the speed $V_+$ in the direction of the electric field $E$. Its zeta potential $\zeta_+$ (or dimensionless mobility) is positive. The anion of negative $\zeta_-$ migrates against the field with the velocity $V_-$.   }
   	\label{fig:sketch}
   \end{figure}

When an  electric field $E$ is applied, an electro-osmotic flow around central anions and cations, referred below simply to as ions, is induced. The electroosmosis takes its  origin in their  EDLs, where a tangential electric field generates a force that sets the fluid in motion. The emergence of this flow in turn provides hydrodynamic
stresses that cause the propulsion of ions with a velocity given by Eq.~\eqref{eq:V_ions} as sketched in Fig.~\ref{fig:sketch}.
If the  clouds of two ions do not overlap, the space between EDLs remains electro-neutral, and, therefore, this region is force-free. Consequently, in this outer region there is no fluid flow. Thus, within our model \emph{the problem of ionic electrophoresis can be treated on a single-particle level} provided the concentration is not too high (below 4 mol/l or so). This, of course, significantly simplifies the problem. However, it would be worthwhile to stress that even the problem for a single colloidal sphere, the so-called ``standard electrokinetic model''~\cite{obrien.rw:1978}, can in general be solved only numerically [since analytical solution of Eq.~\eqref{eq:NLPB} has not been found yet].

In the standard considerations of the electrophoretic migration, an essential supposition is normally
that the radial electric field due to the double layer and the externally applied field may be simply superimposed on each other~\cite{overbeek}.
The P\'{e}clet number is assumed to be small, so that the convective fluxes of ions are ignored. If so, the system of electrokinetic equations is decoupled and the potential obeys the Poisson-Boltzmann equation~\cite{dunweg.b:2008}, implying that the EDLs are undisturbed by the migration.
Such a situation will be referred below simply to as \emph{electrophoresis}. This is certainly not always the case, as the EDL symmetry may be distorted during the migration. This effect that is termed \emph{relaxation}  gives rise to an additional (retarding) field, which is directed oppositely to $E$. Note that (as a second order effect) the relaxation retardation, if any, is smaller than the one, caused by
the motion of the ions of the undisturbed cloud. Since Onsager it is accepted that to arrive at a complete description of migration of ions,  it is necessary  to include this relaxation effect~\cite{onsager.l:1927}.  Note, however, the correction for relaxation goes to zero for large colloidal particles and is never included into consideration~\cite{overbeek}.

\section{Results and discussion}\label{sec:results}

We return to the issue of importance of the relaxation later, by focusing first on electrophoresis solely.  Recall that this  (leading order) contribution to a conductivity is traditionally calculated assuming that the clouds remain undisturbed during the migration.
The electrophoretic mobilities of ions can be defined using the known expression for an electro-osmotic mobility~\cite{smoluchowski.m:1921,vinogradova.oi:2022} taken with the opposite sign
\begin{equation}\label{eq:M}
  M_{\pm} = \dfrac{\varepsilon Z_{\pm}}{4 \pi \eta } =  \dfrac{e }{4\pi \eta 	\ell _{B}}\zeta_{\pm},
\end{equation}
where $Z_{\pm}$ are the electrokinetic or zeta potentials of cations and anions, and $\zeta_{\pm} = e Z_{\pm}/k_B T$.   In such a definition the geometry factor is hidden inside $\zeta$.
Note that by introducing the dimensionless velocity $v_{\pm} = \dfrac{4\pi \eta
\ell _{B}}{e E} V_{\pm}$, one can rewrite \eqref{eq:V_ions} as $v_{\pm} = \zeta_{\pm},$
which points clearly that the electrophoretic speed of ions is set by their zeta potentials.

The expression for zeta potentials of spherical isolating particles, which obey hydrodynamic no-slip boundary condition, has been derived already by~\citet{henry.dc:1931}, and for ions of radius $R_{\pm}$ this can be formulated as
\begin{equation}\label{eq:zeta_general0}
  \zeta_{\pm} =  \phi_{s \pm} + 5 R_{\pm}^{5} \int_{\infty}^{R_{\pm}} \frac{\phi_{\pm}}{ r^{6}} dr  - 2 R_{\pm}^{3} \int_{\infty}^{R_{\pm}} \frac{\phi_{\pm}}{r^{4}} dr.
\end{equation}
In order to take the integrals in Eq.~\eqref{eq:zeta_general0} it is necessary to introduce explicitly the dependence of potentials on $r$. The $\phi_{\pm}$-profiles (and, consequently, $\phi_{s \pm}$) can be computed from Eq.~\eqref{eq:NLPB} leading then to a numerical solution for $\zeta_{\pm}$ of ions.

When the linear approximation of Eq.~\eqref{eq:NLPB} can be applied the local and surface potentials are given, approximately, by
\begin{equation}\label{eq:DH_solution}
 \phi_{\pm} \simeq \pm \dfrac{ \ell_{B} }{ \left(1+\varrho_{\pm}\right)} \dfrac{\mathrm{e}^{\varrho_{\pm} - r/\lambda_D}}{r},  \, \phi_{s \pm} \simeq \pm \dfrac{ \ell_{B} }{ R_{\pm} \left(1+\varrho_{\pm}\right)}.
\end{equation}
It is normally considered that this is justified provided $|\phi_{s \pm}|\leq 1$, but we have recently shown that if radii are small, the linearization of Eq.~\eqref{eq:NLPB} becomes no longer specific just to low potentials~\cite{vinogradova.oi:2023b}. Thus Eqs.~\eqref{eq:DH_solution} should be a sensible approximation for inorganic salt ions of $|\phi_{s \pm}| = O(1)$.

Substituting equation \eqref{eq:DH_solution} into \eqref{eq:zeta_general0} and performing the integration
yields
\begin{equation}\label{eq:zeta_general}
  \zeta_{\pm} \simeq  \phi_{s \pm} \mathcal{F_{\pm}} \simeq \pm \dfrac{\ell_{B}}{R_{\pm}(1+\varrho_{\pm})} \mathcal{F_{\pm}},
\end{equation}
where $\mathcal{F_{\pm}}$ are the Henry special  functions  for nonconducting spherical particles~\cite{henry.dc:1931}
\begin{equation}\label{eq:henry}
  \mathcal{F_{\pm}} = 1 - e^{\varrho_{\pm}} \left[5 \Ei_7(\varrho_{\pm}) - 2 \Ei_5(\varrho_{\pm})\right].
\end{equation}
Here $\Ei_p (\varrho_{\pm})= \varrho_{\pm}^{p-1} \Gamma (1-p,\varrho_{\pm}) = \varrho_{\pm}^{p-1} \int_{\rho}^{\infty} \frac{\mathrm{e}^{-t}}{t^p} dt$ are the generalized exponential integrals. In the  (Smoluchowski) limit of $\rho_{\pm} \to \infty$ the second term in \eqref{eq:henry} vanishes and $\mathcal{F_{\pm}}(\varrho_{\pm} \to \infty) \to 1$. This situation is relevant to large colloid particles, but not to ions of electrolyte solutions, where $\varrho_{\pm} = O(1)$ or smaller. Since $\Ei_p (0) = 1/(p-1)$, it is easy to show that $\mathcal{F_{\pm}}(\varrho_{\pm} \to 0) \to 2/3$. This result is known as the H$\rm{\ddot{u}}$ckel limit~\cite{huckel.e:1924}.

\begin{figure}[h]
	\begin{center}
		\includegraphics[width=0.9\columnwidth]{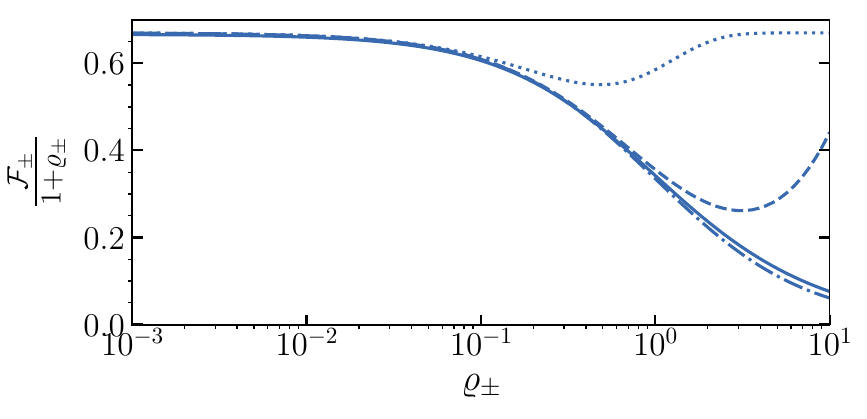}
	\end{center}
	\caption{Function $\mathcal{F_{\pm}}/(1+\rho_{\pm})$ vs $\rho_{\pm}$ with $\mathcal{F_{\pm}}$ calculated from Eqs.~\eqref{eq:henry} (solid curve), \eqref{eq:henry_lin} (dashed curve), and  $\mathcal{F_{\pm}} = 2/3$ (dash-dotted curve). Dotted curve shows the electrophoretic term derived in~\cite{avni.y:2022}.}
	\label{fig:F}
\end{figure}

From Eqs.~\eqref{eq:zeta_general} and \eqref{eq:henry} if follows that the salt dependence of $\zeta_{\pm}$ is hidden only in $\mathcal{F_{\pm}}/(1 + \varrho_{\pm})$. This dependence is illustrated in Fig.~\ref{fig:F}, where we plot $\mathcal{F_{\pm}}/(1 + \varrho_{\pm})$ as a function of $\varrho_{\pm}$. The calculations are first made using $\mathcal{F_{\pm}}$ given by Eq.~\eqref{eq:henry}. It can be seen that the function $\mathcal{F_{\pm}}/(1 + \varrho_{\pm})$ in Eq.~\eqref{eq:zeta_general} decreases [and vanishes in the limit $\lambda_D \to 0$]. Thus on increasing $\varrho_{\pm}$ (reducing $\lambda_D$) the migration speed of ions reduces, which is self-evident since the electrophoresis takes its origin in the EDL.

For small $\varrho_{\pm}$ Eq.~\eqref{eq:henry} can be expanded about $\varrho_{\pm} = 0$ and reduces to
\begin{equation}\label{eq:henry_lin}
  \mathcal{F_{\pm}} \simeq \dfrac{2}{3} \left[1 + \left(\dfrac{\varrho_{\pm}}{4}\right)^2 \right],
\end{equation}
which
has been used to derive the approximate conductivity equation~\cite{vinogradova.oi:2023b}. To examine its range of applicability  the curve for $\mathcal{F_{\pm}}/(1 + \varrho_{\pm})$ calculated with $ \mathcal{F_{\pm}}$ given by  Eq.~\eqref{eq:henry_lin} is included in Fig.~\ref{fig:F}. The accuracy is
quite good for $\varrho_{\pm} \leq 1$, but at larger salt concentrations there is some discrepancy in the direction of larger $\mathcal{F_{\pm}}/(1 + \varrho_{\pm})$ than predicted by using Eq.~\eqref{eq:henry}. Moreover, the function exhibits its minimum at $\varrho_{\pm} \simeq 3$ and then augments, although can only decrease. Note, however, that the electrophoretic term derived by~\citet{avni.y:2022},
\begin{equation}\label{eq:andel}
 \frac{\mathcal{F_{\pm}}}{1 + \varrho_{\pm}} \simeq \frac{2}{3} \left(1-\varrho_{\pm} e^{-a/\lambda_D} \right),
\end{equation}
where $a$ is the is the cut-off length defined as the sum of the cation
and anion crystallographic radii,
fails  at much smaller $\varrho_{\pm}$. The calculation from Eq.~\eqref{eq:andel} is included in Fig.~\ref{fig:F}. It can be seen that it shows significant deviations [from $\mathcal{F_{\pm}}/(1 + \varrho_{\pm})$ obtained using \eqref{eq:henry}] already at $\varrho_{\pm} \geq 2 \times 10^{-1}$ and turns to $2/3$ when $\varrho_{\pm} \geq 2$. This implies that the zeta potentials become the same as at
an infinite dilution, which is counterintuitive.

Also included in Fig.~\ref{fig:F} is the curve calculated using $ \mathcal{F_{\pm}} = 2/3$. It seems clear that this provides $\mathcal{F_{\pm}}/(1 + \varrho_{\pm})$ that practically coincides (when $\varrho_{\pm} \leq 1$) or is very close [at $\varrho_{\pm} = O(1)$] to obtained with Eq.~\eqref{eq:henry}. This implies that the decrease in $|\zeta_{\pm}|$ with salt is to the first order caused only by the reduction of the ion surface potential:
\begin{equation}\label{eq:zeta_ion+2}
  \zeta_{\pm} \simeq \frac{2}{3} \phi_{s \pm} \simeq \pm \dfrac{2 \ell_{B}}{3 R_{\pm}(1+\varrho_{\pm})}.
\end{equation}
The upper bound for the zeta potential is attained when $\rho_{\pm} \to 0$ (an infinite dilution)
\begin{equation}\label{eq:zeta_ion_max}
  \zeta_{\pm} \to  \zeta_{0 \pm} \simeq \pm \dfrac{2\ell_{B}}{3R_{\pm}}.
\end{equation}

\begin{figure}[h]
	\begin{center}
		\includegraphics[width=0.9\columnwidth]{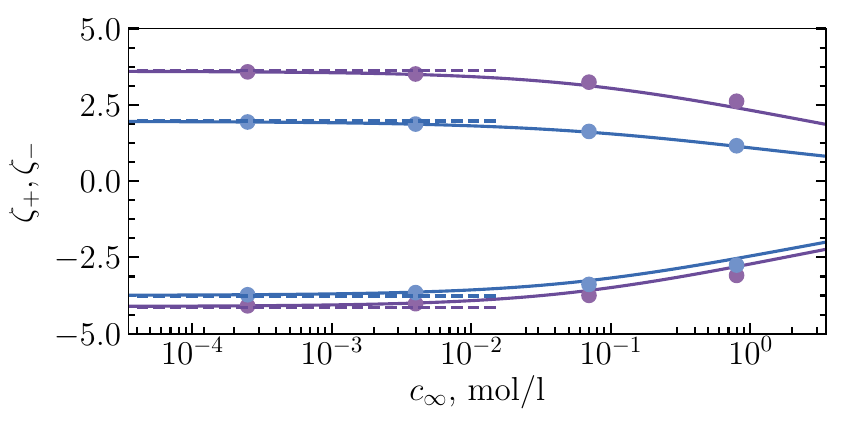}
	\end{center}
	\caption{Zeta potentials $\zeta_{\pm}$ \emph{vs.} $c_{\infty}$ computed for K$^{+}$, Li$^{+}$, Cl$^{-}$, I$^{-}$ (solid curves from top to bottom). Dashed lines are obtained using Eq.~\eqref{eq:zeta_ion_max}. Symbols show calculations from Eq.~\eqref{eq:zeta_ion+2}. }
	\label{fig:zeta}
\end{figure}

Figure~\ref{fig:zeta} includes computed zeta potentials of several ions as a function of salt concentration. For these examples we take from Table~\ref{table:radii}  cations (K$^+$, Li$^+$) and anions (I$^-$, Cl$^-$) of the smallest and largest radius. As a side note, these numerical calculations [based on Eqs.~\eqref{eq:zeta_general0} and \eqref{eq:NLPB}] fully coincide with those made from the first equality in \eqref{eq:zeta_general} using $\phi_{s \pm}$ computed from the nonlinear Poisson-Boltzmann equation and $\mathcal{F_{\pm}}$ given by Eq.~\eqref{eq:henry}.  The straight lines corresponding to $\zeta_{0 \pm}$ given by \eqref{eq:zeta_ion_max} describe perfectly the distinct plateau regions at low  concentrations. On increasing concentration, however, the absolute values of $\zeta_{\pm}$ decrease. Also included are calculations from Eq.~\eqref{eq:zeta_ion+2}. It can be seen that their fit to numerical data is extremely good in the whole concentration range.

\begin{figure}[h]
	\begin{center}
		 \includegraphics[width=0.9\columnwidth]{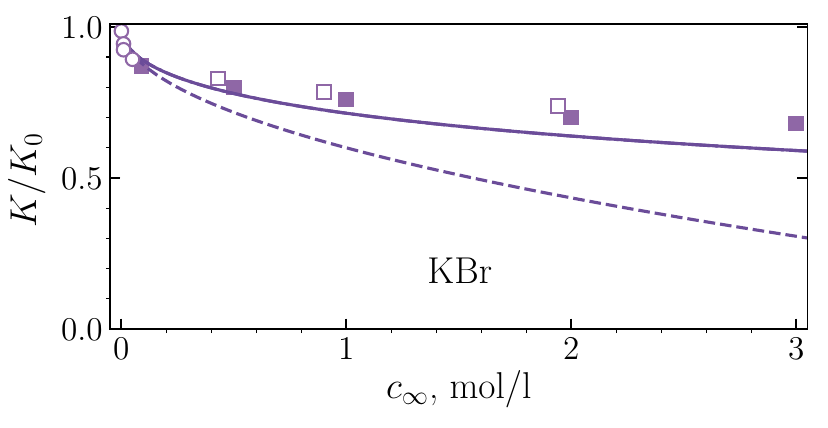} \\
		\includegraphics[width=0.9\columnwidth]{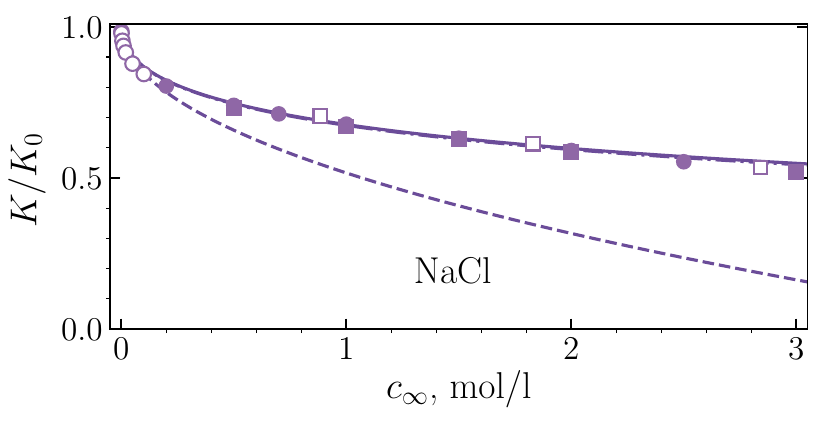} \\
		\includegraphics[width=0.9\columnwidth]{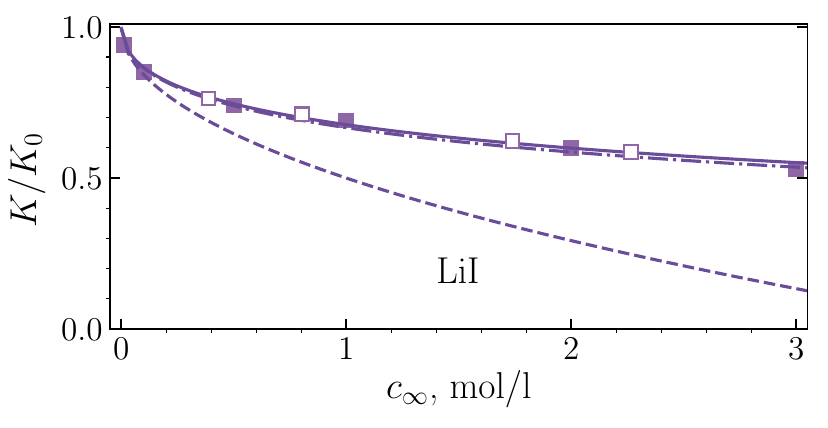}
	\end{center}
	\caption{$K/K_{0}$ as a function of $c_{\infty}$ for KBr, NaCl and LiI aqueous solution calculated from Eqs.~\eqref{eq:MM0}, \eqref{eq:K_appr} and \eqref{eq:onsager} (solid, dash-dotted and dashed curves). Open and filled circles indicate experimental data from \cite{vanysek.r:2000} and \cite{miller.dg:1966}, open and filled squares show data from \cite{dobos.d:1975} and \cite{lobo.vmm:1984}.}
	\label{fig:K_bulk}
\end{figure}

From Eq.~\eqref{eq:M} it follows that
\begin{equation}
M  \simeq \dfrac{e }{4\pi \eta 	\ell _{B}} (\zeta^+ - \zeta^-)
\end{equation}
Substituting \eqref{eq:zeta_ion+2} and performing standard calculations we derive
\begin{equation}\label{eq:MM0}
\dfrac{M}{M_0} \equiv \dfrac{K}{K_0} \simeq  \dfrac{1}{N+1} \left(\dfrac{1}{1+\varrho_+} + \dfrac{N^2}{N+\varrho_+} \right),
\end{equation}
where $M_0$ and $K_0$ are given by Eqs.~\eqref{eq:Kinf}.

Figure~\ref{fig:K_bulk} shows $K/K_0$ calculated from Eq.~\eqref{eq:MM0} as a function of $c_{\infty}$. The calculations are compared with experimental data for three standard salts~\cite{vanysek.r:2000,miller.dg:1966,dobos.d:1975,lobo.vmm:1984}. As seen from Table~\ref{table:radii}, for these examples $N$ varies from 1.1 for KBr, but can be larger than 2 (\emph{e.g.} for LiI). An overall conclusion from this linear scale plot is that the theoretical curve for $K/K_0$ is well consistent with experiment. Only for  KBr  Eq.\eqref{eq:MM0} predicts a slightly lower conductivity at large concentrations,
but still fits data quite well.
This is a startling result since our mean-field theory omits nonidealities that are normally considered to be  important at high salt [and at this point does not include the relaxation effect]. Thus, our results suggest that, if any, they constitute a second-order correction, which could make only a small improvement to the data description.

When $N=1$, we get $\varrho_h = R_h/\lambda_D =  2 \varrho_+/(1+N)= \varrho_+$, and  Eq.~\eqref{eq:MM0} reduces to
\begin{equation}\label{eq:K_appr}
  \dfrac{K}{K_0} \simeq  \dfrac{1}{1+\varrho_h}.
\end{equation}
Eq.~\eqref{eq:K_appr} predicts a monotonic decrease in relative (electrophoretic) conductivity with $\varrho_h$ as it should be, and $K/K_0 \to 0$ as $\varrho_h \to \infty$.  For $\varrho_h \ll 1$  it reduces to the famous Onsager formula for the electrophoretic effect
\begin{equation}\label{eq:onsager}
  \dfrac{K}{K_0} \simeq  1 - \varrho_h.
\end{equation}
The calculation from Eq.~\eqref{eq:K_appr} is also included in Fig.~\ref{fig:K_bulk}. It can be seen that deviations from Eq.~\eqref{eq:MM0} are extremely small. The difference between \eqref{eq:MM0} and \eqref{eq:K_appr}
is always below a few percent. We thus conclude that  compact approximate Eq.~\eqref{eq:K_appr}, which is easy to handle, can safely be employed to interpret data for inorganic salts or for a predictive purpose.

\begin{figure}[h]
\begin{center}
\includegraphics[width=0.9\columnwidth]{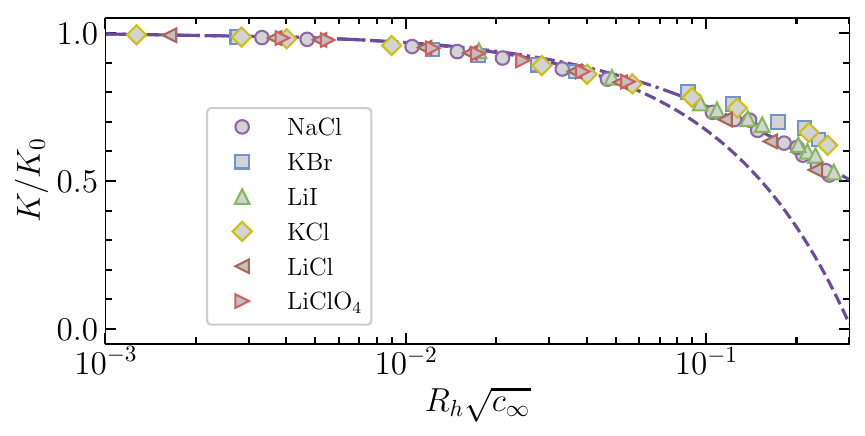}
\end{center}
\caption{$K/K_0$ as a function of $R_h \sqrt{c_{\infty}}$  plotted in a semi-log scale (solid curve) and experimental data for standards salts taken from Refs.~\cite{vanysek.r:2000,dobos.d:1975,lobo.vmm:1984,shedlovsky.t:1932,miller.dg:1966} shown by filled symbols. Dashed-dotted and dashed curve corresponds to calculations from Eqs.~\eqref{eq:K_appr} and~\eqref{eq:onsager}. }
\label{fig:universal}
\end{figure}

It follows from \eqref{eq:DLength} that  $\varrho_h \simeq 3.379 R_h \sqrt{c_{\infty}}$ for a fixed $T = 298$~K, where $R_h$  should be taken in nm.
This implies that the values of $K/K_0$ for any univalent salt solution plotted against (dimensionless) $R_h\sqrt{c_{\infty}}$ should  collapse into a single curve. That this is indeed so is demonstrated in Fig.~\ref{fig:universal}, where we plot in a semi-log scale a relative conductivity calculated from \eqref{eq:K_appr} \emph{vs.} $R_h\sqrt{c_{\infty}}$ along with the experimental data for a variety of inorganic salts at $T = 298~$K~\cite{vanysek.r:2000,dobos.d:1975,lobo.vmm:1984,shedlovsky.t:1932,miller.dg:1966}. Besides salts presented in Table~\ref{table:radii} we include  KCl, LiCl, and LiClO$_4$ using their $R_h$ =  0.127, 0.163, and  0.174 nm, correspondingly, to re-scale the concentration~\cite{kadhim.mj:2020}.
Note that we employ for this plot data for $c_{\infty}$ up to 4 mol/l.
One can see that the approximate theory is in a good agreement with the data for all salts, perhaps with only insignificant discrepancy at intermediate dilution. The discrepancy is always in the direction of slightly smaller $K/K_0$ than predicted by Eq.~\eqref{eq:K_appr}. However, at low dilution our theoretical calculations  practically coincide with the experimental results. Only the data for KBr and KCl show slightly larger $K/K_0$.
Note that there exist some data~\cite{aagren.r:2021} suggesting that conductivity of potassium salts is even higher than in specimen examples used here, but not supported by other experiments.
The  reason  for  qualitative  differences only for these two salts is unclear and is open for discussion, but it must be remembered that this was a first-order calculation only, and we did not expect it to be very accurate.
The curve calculated from Eq.~\eqref{eq:onsager} is also included in Fig.~\ref{fig:universal}. Clearly, the data obtained at high salt are irreconcilable with this linear equation.
\begin{figure}[h]
\begin{center}
\includegraphics[width=0.9\columnwidth]{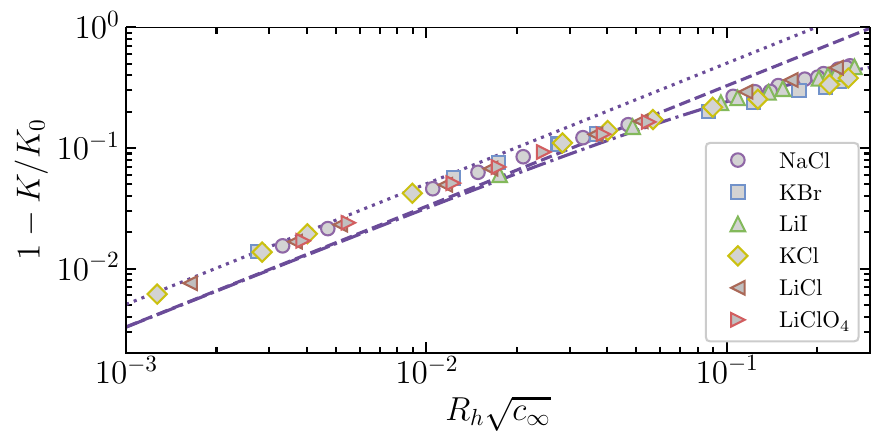}
\end{center}
\caption{A decrement to a relative conductivity, $1 - K/K_0$, plotted against $R_h \sqrt{c_{\infty}}$   in a log-log scale (solid curve). Symbols show the same data as in Fig.~\ref{fig:universal}. Dashed-dotted, dashed and dotted  lines are calculated from Eqs.~\eqref{eq:K_appr}, ~\eqref{eq:onsager} and \eqref{eq:relax}. }
\label{fig:universal2}
\end{figure}

To examine the possible discrepancy from data more closely, the results for salts from Fig.~\ref{fig:universal} are reproduced in Fig.~\ref{fig:universal2}, where a decrement to a relative conductivity, $1 - K/K_0$, is plotted against $R_h \sqrt{c_{\infty}}$ in a log-log scale. It can be seen that in dilute solutions, where the decrease in relative conductivity is so small that was even not discernible on the scale of Fig.~\ref{fig:universal}, the data are slightly above the theory indicating some additional retardation in the ion migration. The discrepancy looks constant at smaller values $R_h \sqrt{c_{\infty}}$, which indicates an increase in the
extra retardation effect with salt, but then the deviations from our (electrophoretic) theory reduce and completely disappear in concentrated solutions.

An explanation can be obtained if we invoke the relaxation effect.
Onsager showed that when $\varrho_h \ll 1$ the electrophoretic and relaxation terms can be considered separately and are summed up~\cite{onsager.l:1927}, which in our notations yields
\begin{equation}\label{eq:relax}
1 -	\dfrac{K}{K_0} \simeq   \varrho_h  + \frac{1}{3}\left( 1 - \frac{1}{\sqrt{2}}\right) \frac{\ell_{B}}{\lambda_D}.
\end{equation}
Indeed, Eq.~\eqref{eq:relax} provides an excellent fit to the data up to $R_h\sqrt{c_{\infty}} \simeq 6 \times 10^{-3}$ (millimolar concentrations) as displayed in Fig.~\ref{fig:universal2}.

Upon  increasing concentration further the experimental data begin to approach the results of calculations from Eq.~\eqref{eq:K_appr}, and in concentrated solutions, $R_h\sqrt{c_{\infty}} \geq 8 \times 10^{-2}$ (or $c_{\infty} \geq 0.5$ mol/l), the theory that includes an electrophoresis solely is in excellent agreement with experiment.
It is natural to conclude that,
 although  in highly dilute solutions the relaxation effect grows  with $c_{\infty}$ [see Eq.~\eqref{eq:relax}], after reaching a maximum this correction to electrophoresis of inorganic salt ions should reduce and disappears at high salt. As the side note, exactly the same trends have been predicted already by~\citet{overbeek} for standard colloid particles.

Experimental observations~\cite{carman.pc:1969,wachter.w:2007}, integral equations~\cite{bernard.o:2023}, and (numerical) calculations based on a nonlocal electrostatics approach~\cite{kalikin.nn:2024} support our conclusion about disappearance of relaxation at high salt. In any event, it seems even intuitively clear that the relaxation, which is caused by  an emergence of an additional retarding field due to a distortion of the cloud symmetry, should vanish at very high salt: no EDLs - nothing to distort, and (as a second order effect) is always smaller than electrophoretic effect. However,  the magnitude of the relaxation term in the approximate conductivity equation derived by~\citet{avni.y:2022} increases monotonically with salt, as highlighted in recent publications~\cite{bernard.o:2023,vinogradova.oi:2024}. When $a = O(\lambda_D)$ this term becomes of the same order as electrophoretic, and in more concentrated solutions even dominates. As a result, when the EDL nearly disappears its relaxation grows. Clearly, it cannot be so, and in essence, the equation by~\citet{avni.y:2022} does not apply at molar concentrations~\cite{note1}.
 However, simple conductivity equation \eqref{eq:K_appr}, which provides a physically correct asymptotic behavior of the electrophoretic term and neglects relaxation, is very well suited to describe concentrated salt solutions.

\section{Concluding remarks}\label{sec:conclusion}

In summary, a physically based simple theory, which strongly appeals to macroscopic arguments, is in quantitative agreement with
existing conductivity data for univalent inorganic salt solutions up to several mol/l or so.  It appears, of course, surprising that compact equation \eqref{eq:K_appr} applies in so large concentration range and for all salts, especially since much more sophisticated microscopic approaches failed to suggest any simple expression of comparable accuracy despite a half-century, if not more, efforts. Recall, however, that it has often been found before that these theories fail when applied to strongly correlated charged fluids (e.g. phase transitions in asymmetric electrolytes), in contrast with the simple  models going back to Debye and H$\rm{\ddot{u}}$ckel~\cite{levin.y:2002}. Our approximate conductivity equation  derived using a simple model for electrolyte solutions and follows from macroscopic treatment appeared to be quite accurate. Some, essentially very small, deviations from data  are in fact irrelevant for most of the applications. Any real system is hardly ideal water-electrolyte solution of a strictly fixed temperature or concentration, and some scatter of the experimental data is always inevitable. Thus the conclusions are unambiguous:
\begin{itemize}
\item The electrophoresis of inorganic ions can be accommodated within  a theoretical framework that employs only a harmonic mean of hydrodynamic radii of ions and
relies on a mean-field description of electrostatics and on a classical approach to colloid electrophoresis. The need to invoke specific constants   as   (arbitrary)   parameters, or to correct  an  ionic concentration  is thereby  removed;
  \item The conductivity of concentrated salt solutions is dictated solely by  ionic  electrophoresis. Although the electrophoretic retardation dominates in highly dilute solutions too, to arrive at a more accurate description of their (in fact, extremely small) conductivity reduction, it seems necessary to include the relaxation correction.

    \end{itemize}

 This  bears  on the whole    question of  what  we mean  by a degree of dissociation that  is often inferred by chemists from the conductivity measurements~\cite{note2}. Our results show that the conductivity is interpreted without invoking the formation of ionic pairs, thus supporting the notion of complete dissociation of  strong  electrolytes. However   remote   from   mainstream  thinking this conclusion  may  seem,  it would be worthwhile to  recall that there are still lingering doubts about the reality of ion pairing, at least for univalent electrolytes in
high permittivity solvents~\cite{marcus.y:2006,levin.y:2002,zavitsas.aa:2001}.

Our considerations can be extended to asymmetric multivalent salts. The same concerns the temperature dependence of $K/K_0$, which follows from our theory, but  requires the validation in terms of fit to experimental results. Another fruitful direction could be to consider the salt-dependence of a mobility of adsorbed ions~\cite{maduar.sr:2015,mouterde.t:2018}, which impacts electrokinetics~\cite{vinogradova.oi:2023}.

On a more fundamental level, it would be interesting to revisit such issues as the diffusion of ions, effective permittivity and viscosity of salt solutions that can be quantitatively discussed using the standpoint taken here. If all these can be interpreted in a different from a common viewpoint way, the implications are large.
A possible result would probably demand a revision of old dogmas and may represent a step forward.

\begin{acknowledgments}

  This research was supported by the Ministry of Science and Higher Education of the Russian Federation. We are
indebted to E.~S.~Asmolov, R.~Buchner, G.~T.~Hefter, B.~Rotenberg, and G.~A.~Tsirlina for  feedback and advice.
\end{acknowledgments}

\section*{DATA AVAILABILITY}

The data that support the findings of this study are available within the
article.

\section*{AUTHOR DECLARATIONS}

The authors have no conflicts to disclose.

%

\end{document}